\documentstyle[prl,aps,twocolumn,epsf]{revtex}

\begin{document}
\draft
\title{Thermodynamics of the Complex  $su(3)$ Toda Theory} 
\author{H. Saleur$^{1,2}$ and B. Wehefritz-Kaufmann$^{1}$}
\address{${}^1$ Department of Physics, University of Southern California,
Los Angeles, CA 90089-0484, USA\\
${}^2$ Laboratoire de Physique Th\'eorique et Hautes Energies, Universit\'e Paris 6,
Paris, France
}
\maketitle
\begin{abstract}
We present the first computation of the thermodynamic properties of the  complex $su(3)$ Toda theory. This is 
possible thanks to a new string hypothesis, which involves 
bound states that are non self-conjugate solutions of the Bethe equations. Our method 
provides equivalently the solution  of the $su(3)$ generalization of the XXZ chain. In the repulsive
regime, we confirm that the scattering theory proposed over the past few years - made only of
solitons with non diagonal $S$ matrices - is complete. But we show that unitarity does not
follow, contrary to early claims, 
 eigenvalues of the monodromy matrix  not being pure phases. In the attractive regime, we find that the 
 proposed minimal solution
of the bootstrap equations is actually far from being complete. We discuss some simple values of the couplings, where, instead of the few conjectured breathers, a very complex structure (involving $E_6$, or two $E_8$) of bound states is necessary to close the bootstrap.

\end{abstract}

\narrowtext

Integrable quantum field theories (QFTs) based on $su(2)$  have been the subject of
intensive studies over the past many years. New theoretical tools like 
the quantum Q-operators \cite{Bazhanov}, the 
Destri De Vega  and Kl\"umper and Pearce equations \cite{DKP}, the connections
with spectral determinant theory \cite{Dorey}, or the relations with elliptic
curves and duality \cite{Fendley} have revealed mathematical structures of remarkable 
depth; they have also made possible the computation of quantities
of experimental interest - and typical of strong interactions - in various
low dimensional condensed matter systems \cite{Fendley}. 

Formal developments, as well as practical applications, would largely benefit from
an extension of these results to the case of other Lie algebras, in particular $su(n)$.
The situation here is somewhat embarrassing, however.  Although the pillars of the $su(2)$ case - 
the $XXZ$ chain and
the associated sine-Gordon model -  have been under control for a long time, even  the
simplest $su(3)$ case is only very partially understood. 

One of the difficulties here - and, from the field theory point of view, one of the most interesting  issues at stake - has to do with  unitarity. Indeed,
the simplest integrable generalizations
of the sine-Gordon model are the  complex \footnote{Real Toda theories involve entirely different issues. For a recent review see \cite{corrigan}.}  affine $su(n)$ Toda theories defined by the Lagrangian:
\begin{equation}
{\cal L}={1\over 2}\left(\partial_\mu\phi^a\right)\left(\partial^\mu\phi^a\right)
-\lambda \sum_{j=0}^n e^{i\beta\alpha_j.\phi}\label{Lagr}
\end{equation}
where $\lambda>0$, $\alpha_1,\ldots,\alpha_{n-1}$ form the root system of the classical lie algebra $a_{n-1}$, $\alpha_0=-\sum_{j=1}^{n-1} \alpha_j$ is the negative of the 
longest root. The conformal weights of  the perturbing operator in
(\ref{Lagr}) are $ \Delta=\bar{\Delta}= {\beta^2\over
4\pi}$. In the following, we shall parameterize $\Delta={t-1\over t}$. The theory described by (\ref{Lagr}) is obviously
{\sl non unitary} at the  classical level.  
The fascinating   possibility was raised
\cite{HOi,GANi} that it could nevertheless describe a unitary field theory in a 
sufficiently  strong quantum regime. This possibility was ruled out in the 
interesting paper \cite{Takacs}, and we confirm and extend their observations 
here.

From a practical point of view, unitarity is not such a key issue. In fact, the most interesting applications of complex Toda theories are potentially found in disordered systems of statistical mechanics, where Toda theories based on superalgebras naturally
seem to appear \cite{andreas,andre}, leading most likely to even stronger 
violations of unitarity. More crucial then are the questions of 
completeness of the bootstrap, the physical meaning of the bound states,
and the calculation of physical quantities. 

The main progress in the study of complex Toda theories have been based on
non perturbative S matrix analysis, following the pioneering work of \cite{HOi}. One of the difficulties in this approach for $su(n)$ is the appearance of 
a large number of poles in the S matrix elements, whose signification is 
not entirely clear: it was argued, after careful analysis of several cases, that most of these poles were not physical, and occurred rather by mechanisms
generalizing Coleman Thun's \cite{COT}.

The issue of the completeness of the bootstrap in \cite{HOi,GANi} could be 
settled by a study of the thermodynamics,
and a computation of the central charge in the UV, 
using the  thermodynamic Bethe 
ansatz (TBA) \cite{ZTBA} \footnote{The approach developed in \cite{DKP} 
bypasses some difficulties in the study of the thermodynamics, but does
not give much information on the spectrum itself \cite{Zinn}.}. However,
 for the imaginary affine Toda theories (or, equivalently,  the corresponding anisotropic quantum spin chains), the TBA
has never been written so far, because of the complexity of the set of solutions: no natural ``string hypothesis'' had 
been proposed, up to now.

In this letter, we present the first solution of this vexing problem  in the 
case of $su(3)$. 
This allows us, in particular, to show  that the quantum theory defined in (\ref{Lagr}) for $n=3$  is
 never unitary, even in the strong quantum regime, and that it presents considerably more bound states than expected.

Our main technical progress  is an understanding of the solutions of the Bethe equations for  systems
based on $su(3)$. Except in the exactly symmetric case, these solutions (the equivalent of the string hypothesis for 
the XXZ chain) had never been found. We hope that their understanding will spur new development in the area: generalizations of the works mentioned in the introduction, as well a calculations of physical properties in the super Toda case, seem particularly timely.

The problem we want to tackle is largely equivalent to the $su(3)$ generalization of the XXZ spin chain. This
integrable chain has been known for a long time \cite{BVV1},and reads
\begin{eqnarray}
H=-\sum_j \left[\sum_{r,s=1,r\neq s}^3 e_j^{rs}e_{j+1}^{sr}+\cos\gamma \sum_{r=1}^3 e_j^{rr}e_{j+1}^{rr}
+\right.\nonumber\\
\left.i\sin\gamma \sum_{r,s=1}^3 \hbox{sign}(r-s)e_j^{rr}e_{j+1}^{ss}\right]
\label{latham}
\end{eqnarray}
Relations of this system with the quantum field theory are {\sl two-fold}. At the physical level, there is 
a simple integrable perturbation - obtained, in the quantum inverse scattering framework, by introducing 
heterogeneities of the spectral parameter, which amount to a staggered interaction - giving rise to (\ref{Lagr}) 
in the continuum limit, as can easily been shown using bosonization, or other arguments \cite{DDV}. In that correspondence, $\gamma={\pi\over t}$. At a more formal level, observe first that the scattering matrices proposed in \cite{Hollowood} 
are {\sl non  diagonal}. To study the thermodynamics of the gas of excitations in the S matrix approach, one needs
to write wave functions, and impose their periodicity. This condition involves ``passing'' a particle 
through a set of other particles with which it scatters non diagonally: the phases obtained in this way 
are, in the inverse quantum scattering framework, eigenvalues of a monodromy matrix \cite{FI}. This monodromy matrix
is nothing but the transfer matrix associated with the hamiltonian  (\ref{latham}) (actually, it  involves a slight generalization
of this hamiltonian with a mixture of the fundamental representation and its 
conjugate), with however a renormalization
of the anisotropy parameter, which becomes  $\gamma={\pi\over {t-1}}$. The case
$\gamma={\pi\over 2}$, for instance (the equivalent of the XX chain, which is not a free problem here however), is a lattice 
regularization of the Toda theory for $\Delta={1\over 2}$; it also turns out in the monodromy problem
for the Toda theory at $\Delta={2\over 3}$.

This double correspondence is very useful. If one is able to solve say the field
theory  problem for a particular value of $\Delta={t-1\over t}$,
one should also be able to diagonalize the lattice model for $\gamma={\pi\over t}$, since it  is just a discretization of this 
quantum field theory. But knowing  the spectrum of (\ref{latham}) for $\gamma={\pi\over t}$ in turn means knowing the spectrum
of the monodromy matrix for this value of $\gamma$. This is nothing but being able to solve the Toda theory for $\Delta={t\over t+1}$! 

Rather than dwell into the lengthy technical details, we would like to discuss first the solution of the lattice model for $\gamma={\pi\over 2}$. 
We consider the slightly more general hamiltonian, where there is an arbitrary mixture of 
the fundamental representation and its conjugate. 
The lattice Bethe equations are \cite{AR}
\begin{eqnarray}
\left({\sinh {1\over 2}(y_j+ i\pi/2)\over\sinh{1\over 2}
(y_j-i\pi/2)} \right)^{N_{\bar{3}}}=&\prod_z {\sinh {1\over 2}
(y_j-z_r-i\pi/2)\over
\sinh {1\over 2} (y_j-z_r+i\pi/2)}\nonumber\\
\left({
\sinh {1\over 2}(z_r + i\pi/2)\over\sinh{1\over 2}
(z_r-i\pi/2)
} \right)^{N_{3}}
=&\prod_y {\sinh {1\over 2}
(z_r-y_j-i\pi/2)\over
\sinh {1\over 2} (z_r-y_j+i\pi/2)},\label{bethe}
\end{eqnarray}
where we have not written a crucial but complicated sign on the right hand sides, and the energy reads $E=-2\sum {1\over \cosh z} -2\sum {1\over \cosh y}$. A combination of numerical studies and  analytical arguments led to the identification of the following sets of roots in
the thermodynamic limit. The $y$ and $z$ can both be real, or both have an imaginary part equal to $\pi$ (antistring). In addition,
it is possible to have complexes, with a two string $z$ centered on an antistring $y$, that is:  $y=r+i\pi,z=r\pm i{\pi\over 2}$ (here
$r$ is a real number, and we identified $r+3i{\pi\over 2}$ and $r-i{\pi\over 2}$), and the same thing with $y$ and $z$ reversed. The existence of these complexes is 
easy to understand. Suppose for instance that $z$ has a positive imaginary part, so the lhs of the second equation in (\ref{bethe})
blows up as $N_{3}\rightarrow\infty$. It is then necessary to have the rhs also blow up, which can be accomplished if there exists a $y$ such 
that $y=z+i{\pi\over 2}$. The same goes if $z$ has a negative imaginary part, so the complexes proposed are the minimal
possible structures leading to real Bethe equations. Together with the real and antistring solutions, they reproduce
the 3+3 degeneracy expected for the fundamental solitons. 
In addition however, another type of  complex (which we call $yz$ in the following) is possible, of the form $z=r+i{3\pi\over 4}$, $y=r+i{5\pi\over 4}$, together with the conjugate.
Usually, one would expect that these two complexes actually come glued together to ensure reality of the Bethe equations and the 
eigenvalues, resulting in 
a sort of ``quartet'' \cite{BVV2}.  This does not seem to be the case here. Rather, to reproduce the correct entropy or
central charge, one needs to treat their densities as {\sl independent}. 

These complexes are the ones which dominate the thermodynamics. As usual, there are many other solutions to the Bethe equations. The existence 
of solutions which are not invariant under complex conjugation being rather unusual, we illustrate it briefly. Figure 1 shows the $yz$ complex obtained by
a numerical solution of (3) for $N_{{\bar{3}}}=0$ and for different values of $N_3$. 
In this example - which seems the simplest possible - the half sum of the imaginary parts of $y$ and $z$ goes to $\pi$ in the thermodynamic limit, but
their difference  does not go to ${\pi\over 2}$. We conjecture that this would however be the case for the overwhelming
majority of such complexes: we checked that only in this case, are the correct scattering theory and 
thermodynamics recovered.

\begin{figure}
\epsfysize=8.5cm
\epsffile{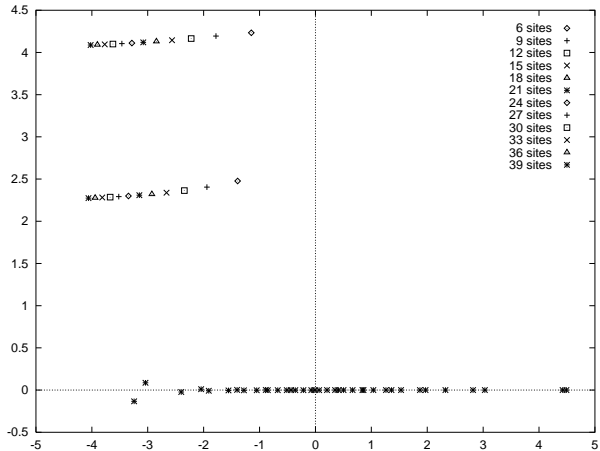}
\vspace{0.5cm}
\caption[]{\label{fig1}
A numerical solution of (3) containing a $yz$ complex for $N_{\bar{3}}=0$
for different values of $N_3=6,9,\cdots,39$. For $39$ sites we show all roots
belonging to this particular state, whereas for all other lattice lengths
only the $yz$ complex is shown. }
\end{figure}

The lattice model without heterogeneities has a continuum limit which is a conformal field theory, made of two 
bosons compactified on a triangular lattice. The latter is unitary, and we have checked numerically that the imaginary parts 
of the eigenenergies of (\ref{latham}) scale to zero faster than their real parts, ensuring reality of the 
conformal weights, and, actually, unitarity of the conformal field theory, 
despite the non-hermicity of the hamiltonian. Meanwhile, the lattice model
with  heterogeneities does  have a continuum limit described by a complex Toda theory, with $\Delta={1\over 2}$. In this case,
the complexes correspond to physical particles, the bound states discussed in \cite{HOi,GANi}. The fact that they are described 
by solutions of the Bethe ansatz which are not self conjugate corresponds to the shocking fact that, in the quantum
field theory, their S matrix is not a pure phase (and ``naive'' unitarity, in the sense of the scattering theory, is broken). 

Accordingly, a  crucial feature related with the $yz$ complexes is that they give rise to non real kernels in the continuum Bethe ansatz. For instance,
the kernel for the scattering of a real $z$ through a $yz$  complex $z=r+i{3\pi\over 4}$, $y=r+i{5\pi\over 4}$
is  ${1\over i}{d\over dz} \ln\Phi=-1/\cosh\left(z-r+{i\pi\over 4}\right)$. Although at equilibrium the densities of the two 
types of conjugate complexes are equal, it is necessary to let them vary independently to get the correct entropy. The physical results will
be the same as the ones that would be obtained with a different theory, where the scattering kernels would be real, and given by
the real part of the true kernels. This is what we will consider in the following, to make the notations simpler.

The lattice model whose equations are (\ref{bethe}) occurs also in the problem of diagonalizing the monodromy matrix for $\Delta={2\over 3}$. The complexes
just discussed have no physical meaning then: they just correspond to pseudo particles of zero mass. However, the fact that they
are not self conjugate means that eigenvalues of the monodromy matrix will not, in general, have modulus one. Contrary to early claims \cite{HOi}, unitarity 
is strongly violated in the theory, even in the absence of bound states.
The physical implications of having ``monodromies'' which are not pure phases 
are not clear to us: presumably $S$ matrix elements have to be considered as 
formal objects used to build wave functions, and cannot be given a reasonably meaning in terms of scattering processes \footnote{A somewhat related problem occurs for instance in the massless description of conformal field theories.}.   

This basic structure generalizes easily to the case $t$ integer. For $\gamma={\pi\over t}$, the Bethe equations contain in addition a term of interaction between $y$ roots, and between $z$ roots. The solutions
are the usual $y$ or $z$ real, antistrings, or $2,\ldots,t-1$ strings, plus 
$y$-$t$ strings centered on $z$ antistrings, and the same thing
with $y$ and $z$ reversed. In addition, the $yz$ complexes
now are of the form $z=r+i\pi -i {\pi\over 2t}$, $y=r+i\pi+ i{\pi\over 2t}$, and the same with $y$ and $z$ exchanged.

To proceed, one has to use these results to solve the monodromy problem. At $\Delta={t-1\over t}$, passing a soliton through a set of $N_3, N_{\bar{3}}$ solitons at various rapidities gives rise 
to a```phase'' which is an eigenvalue of the monodromy matrix. The latter is essentially the transfer matrix associated with (\ref{latham}),
but for $\gamma={\pi\over t-1}$. Its eigenvalues are obtained by generalizing slightly the Bethe ansatz equations, so the arguments in the lhs of (\ref{bethe}) are shifted by the corresponding rapidities, and using the  generalized string hypothesis described previously.

A few technical details are involved, which are completely equivalent to what happens in the $su(2)$ case \cite{FS}. We will
only describe the end result, which  is quite simple. For the case $t\rightarrow\infty$, the TBA has been known for a long time  \cite{Nathan,Ravanini}, and has a structure that
mimics the known one for $su(2)$, with an infinity of massless nodes corresponding to the usual strings of the Bethe equations. Like for
the $su(2)$ case, the introduction of anisotropy {\sl truncates} this to a finite number of strings: the truncation
has to be completed by the proper ``end structure'' of the diagram. Like in the $su(2)$ case, this structure, for $su(3)$, is given 
by  the Bethe roots for $\gamma={\pi\over 2}$. After some manipulations, the results are the following. 
We call $\sigma,\sigma^h$  the densities of the solitons at rapidity $\theta$, and $m$ the mass of the fundamental soliton. Then 
\begin{eqnarray}
\sigma^3+\sigma^{3 h}=&m\cosh\theta+ \phi\star \left(\rho_1^{3 h}+
\sigma^{\bar{3}h}\right)\nonumber\\
\ldots \nonumber\\
\rho_n^{3}+\rho_n^{3 h}=&\phi\star \left(\rho_{n-1}^{3  h}+\rho_{n+1}^{3  h}+\rho_n^{\bar{3}}\right)
\ldots 
\end{eqnarray}
and similarly for $\bar{3}$, with $\rho_0\equiv\sigma$. We recognize here the
standard equations for the minimal model \cite{Hollowood}. In addition, in this untruncated case, we need the ``closure'' relations, which are the key to the whole problem. They read here
\begin{eqnarray}
\rho_{t-3}^{3}+\rho_{t-3}^{3 h}&=&\phi\star \left(\rho_{t-4}^{3 h}+\rho_{t-2}^{3 h}+\rho_{t-3}^{\bar{3}}+
\rho_{a_1}^{3}
+\rho_{a_2}^{3}\right)\nonumber\\
&+&\psi\star \left(\rho_b^{3}+\rho_b^{\bar{3}}\right)\nonumber\\
\rho_{t-2}^{3}+\rho_{t-2}^{3 h}&=&\phi\star \left( \rho_{t-3}^{3 h}+\rho_{t-2}^{\bar{3}}-\rho_{a_1}^{\bar{3}}-\rho_{a_2}^{\bar{3}}\right)
\nonumber\\
&-&\psi\star \left(\rho_b^{3}+\rho_b^{\bar{3}}\right)\nonumber\\
\rho_{a_i}^{3}+\rho_{a_i}^{3 h}&=&\rho_{t-2}^{3}+\rho_{t-2}^{3 h}
\end{eqnarray}
and similarly for $\bar{3}$; finally
\begin{eqnarray}
 \rho_b^{3}+\rho_b^{3 h}&=&\psi\star \left(\rho_{t-3}^{3 h}+\rho_{t-2}^{3}-
\rho_{a_1}^{3}-\rho_{a_2}^{3}+3\to \bar{3}\right) \nonumber\\
&-&\phi\star \left(\rho_b^{3}+\rho_b^{\bar{3}}\right)
\end{eqnarray}
In these equations, the kernels are defined by their fourier transforms \footnote{ We define $\hat{f}(x)={1\over 2\pi}\int f(\theta) e^{-inx\theta/\pi}d\theta$.}
$\hat{\phi}={1\over 2\cosh x}$ and $\hat{\psi}={\cosh x/2\over 2\cosh x}$. The subscripts $a_i$, $i=1,2$ stand for antistrings or $t$ strings centered on antistrings; $b$ stands for $yz$ complexes.

We set $e^{-\epsilon_0/T}=\sigma/\sigma^h$, $e^{-\epsilon_n/T}=\rho_i^h/\rho_i$, $n=1,\ldots,t-2$, $e^{-\epsilon_a/T}=\rho_{a_i}/\rho_{a_i}^h$,
$e^{-\epsilon_b/T}=\rho_b/\rho_b^h$ (color labels are kept implicit here). Introducing the usual variables $x_\alpha=e^{-\epsilon_\alpha/T}$, the  TBA in the UV limit reduces to the system ($x_{t-2}=1/x_a$):
\begin{eqnarray}
x_0=&(1+x_1)^{1/2} \left(1+{1\over x_0}\right)^{-1/2}\nonumber\\
\ldots &\nonumber\\
x_n=&(1+x_{n-1})^{1/2}(1+x_{n+1})^{1/2}\left(1+{1\over x_n}\right)^{-1/2}\nonumber\\
\ldots &\nonumber\\
x_{t-3}=&(1+ x_{t-4})^{1/2} (1+x_a)^{3/2} \left(1+{1\over x_{t-3}}\right)^{-1/2} (1+x_b)\nonumber\\
x_a=&(1+x_{t-3})^{1/2} \left(1+{1\over x_a}\right)^{-1/2} (1+x_a)^{-1} (1+x_b)^{-1}\nonumber\\
x_b=&(1+ x_{t-3})\left(1+{1\over x_a}\right)^{-1} (1+x_a)^{-2} (1+x_b)^{-1}
\end{eqnarray}
The solution  of this system is
\begin{eqnarray}
 x_j=&{ (j+1)(j+4)\over 2}&,\ \ j=0,\ldots,t-3\nonumber\\
x_a=&{t-1\over t+1}&\nonumber\\
x_b=&{(t-1)^2\over 4t}&\
\end{eqnarray}
In the IR limit, we get an identical TBA, but with the replacement $t\to t-1$ exactly as in the $SU(2)$ case \cite{FS},
because the first two nodes become infinitely massive, $\epsilon_0=\infty$. The whole TBA system is actually quite 
similar to the one for the sine-Gordon model: the ``left part'' can be conveniently encoded in a ladder diagram 
with base the Dynkin diagram of $a_2$, and in the isotropic limit $t\to\infty$, this is all that matters. For finite $t$ however,
the diagram has to be closed to the right, and the closing terms are more complicated than for sine-Gordon: in addition
to 3+3  nodes standing algebraically for the representations $3$ and $\bar{3}$ (in the sine-Gordon case, there are two such nodes
corresponding to the self-conjugate fundamental representation), the closure requires the nodes associated with the non real 
solutions of the Bethe  ansatz, whose algebraic meaning has eluded us up to now.
The TBA diagrams for SU(2) and SU(3) are shown in Fig.\ 2.

\bigskip
\noindent
\centerline{\hbox{\rlap{\raise28pt\hbox{$\hskip4.5cm\bigcirc\hskip.25cm t-2$}}
\rlap{\lower27pt\hbox{$\hskip4.4cm\bigcirc\hskip.3cm a$}}
\rlap{\raise15pt\hbox{$\hskip4.1cm\Big/$}}
\rlap{\lower14pt\hbox{$\hskip4.0cm\Big\backslash$}}
\rlap{\raise13pt\hbox{$\hskip 0.2cm 0\hskip0.9cm 1\hskip0.9cm 2$}}
$\bigotimes$------$\bigcirc$------$\bigcirc$-- -- -- 
--$\bigcirc\hskip.5cm t-3 \hskip1cm \Delta=\frac{t-1}{t}$ }}

\bigskip
\begin{figure}
\epsfysize=4.0cm
\epsffile{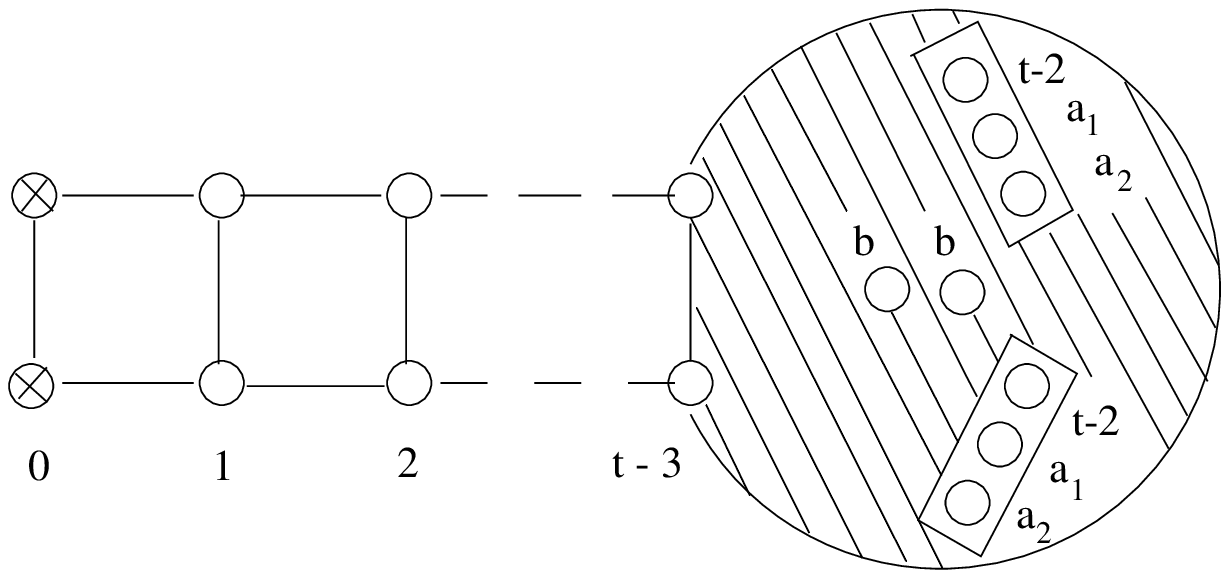}
\vspace{0.6cm}
\caption[]{\label{fig2}
The TBA diagrams for $SU(2)$ and $SU(3)$}
\end{figure}

Using the basic  identity \cite{K}: $L(x)+L(y)=L(xy)+L\left[{x(1-y)\over 1-xy}\right]+L\left[{y(1-x)\over 1-xy}\right]$, one finds
\begin{eqnarray}
\sum_{j=0}^{t-3} 2 L\left[{j^2+5j+4\over j^2+5j+6}\right] +6 L\left[{t-1\over 2t}\right]
\nonumber\\
+2 L\left[\left({t-1\over t+1}\right)^2\right]=(t-1)~~ {\pi^2\over 3},
\end{eqnarray}
and therefore,  $c= 2(t-1) - 2(t-2)=2$, as required. Note that for the sine-Gordon case, the sum of dilogarithms for 
the UV (IR)  diagram was  $(t-1) {\pi^2\over 6}$ (resp. $t-2$) instead, generalizing nicely the combinatorics to the case of rank two. 

Still like in the sine-Gordon case, the same TBA diagram supports the scattering theory for generalized supersymmetric 
extensions: by putting the mass term on the $k^{th}$ node, the central charge now becomes $c={8k\over k+3}$,
in agreement with the central charge for $SU(3)_k$ theories. 

The Toda theory
 can be twisted by coupling a charge to the total 
diagonal $U(1)$ current. The kinks come in the form of $3$ and $\bar{3}$, and the total charge can be recast in terms
of the solutions of the associated, non diagonal scattering problem, as 
$Q\propto (t-1) \int \rho^3_{t-2}-\rho_{a_1}^{3}+(3\to\bar{3})$. The thermodynamics is more complicated to carry out than in the sine-Gordon case however, densities of holes and particles mixing up in a complex fashion in the right hand side of the universal Bethe equations. It can still be shown  that, with a fugacity $-1$ for the solitons, the diagram truncates, leaving only the nodes $j=0,\ldots, t-5$, for $t\geq 5$, the well known result for the minimal model
$SU(3)_1\times SU(3)_{t-4}/SU(3)_{t-3}$. 
Note that more nodes
disappear than in the sine-Gordon case: the case where the minimal theory has $c=0$  now corresponds to a dimension $\Delta={3\over 4}$ in the Toda theory. 

The study of the points $\Delta={t-1\over t}$, $t$ integer, therefore presents few surprises,
except maybe the non unitarity, due to the complex solutions of the Bethe 
equations. We do not expect a qualitatively different behaviour when $t$ is not an integer,
provided $\Delta$ lies in the repulsive regime, $\Delta\geq 3/5$, 
where no bound
states are expected. In contrast, the attractive regime turns out to be considerably more complicated. 

From a technical point of view, the points $\Delta={1\over t}$, could be studied as the continuum limit of the chain with $\gamma={\pi\over t}$ still, but
with  the opposite sign of the hamiltonian. This observation is crucial in the case of the XXZ chain, and allows one to essentially use the same type of solutions of the Bethe equations to study both the attractive and the repulsive regime: it is 
well known that the TBAs for these two cases are very closely related. In the $su(3)$ case, the situation is more involved: depending on the sign of the hamiltonian, it is not the same type of solutions  of the Bethe equations that determines the thermodynamics. This can be illustrated in the case $\gamma={\pi\over 2}$. Taking say a chain made only of $3$'s, the ground state
is $y$ and $z$ real for one sign, $y$ antistring and $z$ real for the other. This still is like for the XX chain. However, the other solutions of the Bethe equations are different: for instance, instead of having two string $z$ centered on antistrings $y$'s, one has now two string $z$ centered on real $y$'s. Things get worse as $t$ increases, and solutions that were possible but did not contribute to the thermodynamics for one sign of the coupling, start being important for the other.

The net result is that we have not been able to reproduce the simple $c=2$ result for any point in the attractive regime except the $\Delta={1\over 2}$ case. To illustrate the nature of the difficulties, let us concentrate for a while on the theories for which $\lambda={4\pi\over\beta^2}-1$ is of the form $\lambda={2n+1\over 3}$, or $\lambda={2n+2\over 3}$. In that case, the deformation parameter for the quantum group symmetry of the scattering theory, $q=-\exp(i\pi\lambda)$ is a cubic root of unity, and the q-dimension of the $3$ or the $\bar{3}$ vanish exactly, $(3)_q=0$. This means that the RSOS truncation \cite{SR,BL} of the theory is very simple: the kink part entirely disappears from the scattering, and one is left with a diagonal scattering for the scalar bound states. This scattering leads to straightforward TBA calculations,
which one can compare against the predictions of conformal field theory. The
disagreement with the scattering theory proposed in \cite{GANi} is  considerable. Let us give two examples. For $\lambda={4\over 3}$, the analysis of \cite{GANi}
predicts only a pair of bound states: to reproduce the effective central charge
$c_{\hbox{eff}}=6/7$ together with its apparent conformal weight $\Delta_{\hbox{eff}}=1/7$,
one needs in fact 6 bound states, essentially reproducing the well known $E_6$ scattering theory \cite{esix}. Note that in this theory, there are pairs of conjugate particles, but also self-conjugate ones: the issue of whether bound states in $su(3)$ should always appear in pairs is not clear, and this counter example demonstrates it is not always the case. 

If the appearance of $E_6$ is somewhat surprising, an even bigger surprise is encountered  for $\lambda={5\over 3}$. In this case, the effective central charge is $c_{\hbox{eff}}=1$, and the apparent weight $\Delta_{\hbox{eff}}=1/16$. The scattering 
theory in \cite{GANi} predicts two pairs of bound states, i.\ e.\  $4$ scalar particles. To close the bootstrap, it turns out that one needs in fact  $16$ particles,
with a scattering theory that is a sort of double $E_8$\cite{eeight}: if one labels the $E_8$ particles by $j$, the $su(3)$ particles by $B_j$, then one has the identity 
$S_{B_j\bar{B}_j}S_{B_k\bar{B}_k}=S_{jk}$:
this guarantees that the central charge in the UV is $2\times {1\over 2}=1$\footnote{This doubling is similar to what happens in the scattering theory for the 3 state Potts model, with $c=4/5$, compared with the scattering theory for the Yang Lee model, with $c_{\hbox{eff}}=2/5$.}, as required. 

The ratio of the actual number of bound states and the minimal one
is actually growing very quickly, and it is not clear what to do to close
the bootstrap even for the simplest non trivial reflectionless point,
$\Delta={1\over 3}$, where the conjectured scattering theory
\cite{GANi,HOii}, involving the 
fundamental solitons $A_0,\bar{A_0}$, excited solitons $A_1,\bar{A}_1$, and breathers $B_1,\bar{B}_1$, $B_2,\bar{B}_2$, gives only $c=1.90821$. Note that this point is quite special, since the particles 
can be gathered into 2 eightuplets of particles of mass $M$ and $\sqrt{3}M$, indicating the presence of a hidden 
$D_4^{(3)}$ symmetry \cite{V,TA}. Introducing more particles increases $c$, but 
we haven't found a closure of the bootstrap, even after introducing large numbers of bound states \footnote{Let us stress here that the possibility of having missed some 
CDD factors in the original scatterings should be considered as excluded, since our lattice calculations (see also \cite{Rafael}) do fully reproduce the accepted kernels. }.

The repulsive regime for $su(n)$, $n\geq 4$, presents similar difficulties: the 
scattering theory of \cite{GANi,HOii} is incomplete even at the point $\Delta={1\over 2}$, and there again, more bound states have to be introduced. 

In conclusion, except for the simplest case of $su(3)$ in the repulsive regime,
the scattering theory of the complex Toda theories is not under control, in our opinion. The large body of literature on the subject (see e.\ g.\  \cite{refs} 
for 
some recent developments)  has identified a minimal structure for it, 
but reality is more complicated (a fact well known to some experts). 
How complicated 
is not clear to us:  without an algebraic understanding of 
the bound states, tackling the general case is a very laborious task, and it is not even clear that the bootstrap will 
close  with a finite number of bound states, except for some special values: clearly, more work is needed in this direction.

Meanwhile, the solution of the $su(3)$ theory in the repulsive regime
opens the way to several problems of physical interest, in particular study of  
$su(3)$ anisotropic Kondo problems, and $su(3)$ tunneling problems, on which we hope to report soon. 

\bigskip
\noindent{\bf Acknowledgments}: we thank P. Fendley for very useful discussions
and G. Takacs for comments on the manuscript.
The work was supported by the DOE and the NSF (under the NYI program). 
B.\ W.-K.\ acknowledges support from the 
Deutsche Forschungsgemeinschaft (DFG) under the contract KA 1574/1-1.

\end{document}